\documentstyle[prl,aps,preprint]{revtex}
\begin{document}
\draft
\preprint{CTP-TAMU 21/94}

\title{Self-diffusion in random-tiling quasicrystals}

\author{Marko Vukobrat Jari\'c}

\address{Department of Physics, Texas A\&M University,
College Station, TX 77843, USA}

\author{Erik S. S\o rensen}

\address{University of British Columbia, Vancouver, BC, V6T 1Z1, Canada}

\maketitle
\begin{abstract}
The first explicit realization of the conjecture that
phason dynamics leads to self-diffusion in quasicrystals is
presented for the icosahedral Ammann tilings.
On short time scales, the transport is found to be subdiffusive
with the exponent $\beta\approx0.57(1)$, while on long time
scales it is consistent with normal diffusion that is
up to an order
of magnitude larger than in the typical room temperature vacancy-assisted
self-diffusion.
No simple finite-size scaling is found, suggesting anomalous corrections
to normal diffusion, or existence of at least two independent
length scales.
A connection with transport in membranes is also noted.
\end{abstract}

In addition to the uniform translations, quasiperiodic (incommensurate)
crystals possess
another Goldstone mode that is associated with relative translations
between incommensurate components of density waves.\cite{bak1,kkl}
This continuous zero-energy mode, the
uniform phason displacement, does not generally correspond to a
continuous, zero-energy path in the configuration space of the
quasiperiodic crystal.
{}For quasiperiodic structures with certain noncrystallographic symmetries
(quasicrystals), a uniform
phason displacement is necessarily mapped on a discrete rearrangement of
particles in the configuration space.\cite{bak2,frenkel}
These rearrangements, which are called phason flips
(see Fig. 1b),
have interesting topological properties\cite{frenkel,katz,kalugin}
that suggest a novel mechanism of mass transport
in quasicrystals.\cite{kk}
In particular, it was shown in Ref. \cite{frenkel} that it is possible
to make a closed path
in the phason displacement space of a quasicrystal, such that
starting from a quasicrystal structure in the real space one ends up
with the identical structure except for a permutation of its atoms
(see Fig. 1c).

These observations raise interesting questions regarding the
mass transport in quasicrystals.
{}First, can an atom be moved an arbitrarily large but finite
distance in a quasicrystal by combining a finite number of phason
flips?
If the answer to this question is yes, then, what kind of transport
results from the phason dynamics?
By considering a perfectly quasiperiodic quasicrystal as a
cut through a higher-dimensional periodic
crystal (hypercrystal),\cite{bak1,kkl}
it was argued in \cite{kk}
that the answer to the first question is affirmative for nearly
perfect quasicrystals whenever the
hyperatoms of the associated hypercrystal are sufficiently extended.
This was explicitly demonstrated for
the transport of vertices in a nearly perfect octagonal tiling.\cite{kk}
A similar proof also exists for decagonal (Penrose) and icosahedral
(Ammann) tilings.\cite{kprivate}
{}Furthermore, it was conjectured in \cite{kk} that the phason dynamics
would lead to a diffusive transport with a characteristic Arrhenius
temperature dependence of the associated diffusion constant that
crosses-over from one characteristic activation energy at low temperatures
to a smaller activation energy at high temperatures.

The main motivation for this letter is to prove the high temperature
limit of the above conjecture for quasiperiodic icosahedral Ammann
tilings and their periodic approximants.
Their equilibrium properties have been extensively studied using the
Monte Carlo phason dynamics,\cite{tang,shaw}\ which we also employ
here.
By measuring the mean square displacement of vertices $R^2$ as a function
of time $t$ and size $L$ of periodic approximants of the Ammann tilings,
shown in Fig. 2, we show that the long-time transport
in the infinite Ammann tilings is diffusive, $R^2=0.0012(1)~t$,
with $R$ in units of the tile edge length $l$ and $t$ in units of the
Monte Carlo sweeps (MCS).
When translated into physical units, this gives a diffusion
constant that can be an order of magnitude larger than the one obtained from
vacancy-assisted mechanism at room temperature.
This suggests that the phason flips might be the
dominant mechanism of self-diffusion in quasicrystals, at least at
temperatures where the effective energy for the flip creation
begins to saturate.\cite{kk}
We also find a universal, size-independent, subdiffusive short-time
behavior, $R^2\sim t^{0.57(1)}$.

We find a similar behavior for infinite periodic approximants
of the Ammann tiling.
Analyzing transport data shown in
{}Fig. 2 for
different finite periodic approximant sizes, we find that the
data is not consistent with a simple finite-size scaling.
This indicates a possible presence of corrections to scaling,
existence of at least two independent
length-scales (or physical mechanisms) governing the phason-assisted
self-diffusion, or anomalous long-time correlations.

We consider the ensemble of three dimensional
random icosahedral Ammann tilings.\cite{tang}
Each member of the ensemble is a tiling by prolate and oblate
rhombohedra with edges parallel to the six five-fold
symmetry axes of the icosahedral symmetry as shown in Fig. 1a.
{}Furthermore, the orientational icosahedral symmetry of each tiling
is guaranteed by the fact that prolate or oblate rhombohedra
of all icosahedrally equivalent orientations appear with equal density,
with the ratio of the overall densities of the prolates and oblates equal
to $\phi=(1+\sqrt{5})/2$.
Starting from the perfect quasiperiodic Ammann tiling, members of the
ensemble are constructed by successive phason flips (rearrangements of the
rhombic dodecahedra) like the one shown in Fig. 1b.
In equilibrium, these phason flips do not destroy
long range order (quasiperiodicity) of
the perfect Ammann tiling.\cite{tang,shaw}

We employ the discrete Monte Carlo dynamics of phason flips,
that was also used to equilibrate Ammann tilings.\cite{tang}
The random Ammann tiling is considered a plausible prototype of
icosahedral quasicrystals.
Moreover, several structure models of icosahedral or decagonal quasicrystals
are based on atomic decorations of Ammann\cite{review} or Penrose\cite{burkov}
tilings.
In addition, although the real dynamics of atoms is certainly not Monte Carlo,
the phason flip in a real crystal is likely to correspond to a movement
of an atom (or a group of atoms) from one  potential energy well to
another, crossing an energy barrier.
Thus, the ``real" dynamics could be mapped to the Monte Carlo
dynamics by relating the ``real" time $t$ to the Monte Carlo time $t_{MC}$ by
$t_{MC}=t \nu\exp(-\epsilon/k_B T),$
where $\nu$ is the attempt frequency for crossing the barrier and
$\epsilon$ is the barrier energy.

Under the phason dynamics, vertices ${\bf r}_i$ of a tiling perform walks that
we characterize as phason-assisted self-diffusion.
The diffusion can be quantified in the usual way, in terms of the time
dependence of the average square end-to-end distance of the walks
(i. e. of the displacements of vertices),
\begin{equation}
\label{eq1}
R^2={1\over N}\sum_{i=1}^N|{\bf r}_i(t)-{\bf r}_i(0)|^2,
\end{equation}
where $N$ is the number of vertices in the tiling.
In addition to the time dependence of $R^2$, we are also interested
in its dependence on the linear tiling size $L\sim N^{1\over3}$.
We start from periodic approximants of the Ammann
tiling,\cite{tang,shaw,jq}
the infinite  periodic tilings related to the perfect
Ammann tiling by approximating $\phi$ with its truncated
continued fraction expansion, $\phi\approx F_{n}/F_{n-1}$,
where $F_n$ is a Fibonacci number, $F_{n+2}=F_{n+1}+F_n$, $F_0=F_1=1$.
These are cubic tilings with $N_n=4(2F_n^3+3(F_n+F_{n-1})F_nF_{n-1})$
vertices per cubic unit cell of edge length
$L_n=2l(F_n\phi+F_{n-1})/\sqrt{\phi+2}$.
Then, for an approximant, we carry out the phason dynamics
with periodic boundary conditions applied to a cubic unit cell
of size $L=m L_n$ with $N=m^3 N_n$ vertices.
In this way, not only can we investigate the infinite
quasicrystal as $n\rightarrow\infty$, but for a given $n$
we can also investigate infinite crystals as $m\rightarrow\infty$.

The Monte Carlo phason dynamics is implemented in the following way.
A vertex is selected at random and if it is of the type shown in Fig. 1b,
it is flipped.
Otherwise, the procedure is repeated.
Since each vertex retains its identity in a single flip, and the
dynamics consists of a sequence of isolated flips, it is possible
to follow a single vertex and measure its position ${\bf r}(t)$ as a
function of time.
The unit of time is one Monte Carlo sweep, equal to $N$
flip attempts.
It should be noted that since we are interested here in the high
temperature limit, we do not invoke any energetics in the flips.

Starting from a perfect tiling we first equilibrate it for
$N$ MCS (the characteristic relaxation time for equilibration
with the phason flip dynamics was found\cite{tang,shaw} to
scale with the system size as $L^2\sim N^{2/3}$).
{}Following this equilibration period, we record the position
of each vertex ${\bf r}_i(t)$ for the ensuing $2\times10^5$ MCS at intervals
of 5 MCS at short times and 1000 MCS at long times.
We investigated tilings with $n$ up to 7 and $m$ up to 4.
{}For all cubic tilings, periodic boundary
conditions that respect the cubic symmetry of the tiling
can be imposed on simple cubic (sc), face centered cubic (fcc), and
body centered cubic (bcc) unit cells that contain
the primitive unit cell of the tiling.
{}For each periodic tiling (i.e., each $n$), this leads to three
families of periodic boundary condition unit cells with $N=m^3N_n$,
with $m$ equal to an integer, $2^{1\over3}$ times an integer,
or to $4^{1\over3}$ times an integer, for sc, fcc, or bcc boundary
conditions, respectively.
Tilings with $n$=1, 4, 7, ... (that is, those where {\it both} $F_{n}$ and
$F_{n-1}$ are odd) are bcc, so that $m$ takes the value $2^{-{1\over3}}$
times an integer for the bcc boundary condition family.
All other tilings are sc.
In order to estimate the error bars, we repeat
the simulation several times for each tiling ranging from 400 times,
for $n$=2, to 2 times, for $n=7$.
It can be shown that for $n=1$ no vertex can be transported
beyond a single flip and, therefore, we consider only $n>1$.

We show in Fig. 2 the mean square displacement averaged
over all vertices and samples as a function of time for all
tilings with $n$=2,3,4,5,6, $m$=1, and $n=7$, $m=2^{-{1\over3}}$.
The short time behavior is independent of the system size $L$ and can be
fitted to the power-law dependence $R^2=at^\beta$ with
$\beta=0.57(1)$, and $a=0.0310(1)l^2$/MCS$^\beta$.
Independent of the periodic unit cell size,
the long time behavior is consistent with a {\it linear} dependence
of $R^2$ on time $t$.
We determine the diffusion constant, $D$, by fitting $R^2(t)$ between
$t=10^5$ MCS and $2\times 10^5$ MCS to the ``normal" diffusion form
\begin{equation}
R^2(t)=D_Lt+C_L.
\label{normald}
\end{equation}
$D$ depends on $L$, as shown in the inset in Fig. 2, and we
extrapolate to the $L\rightarrow\infty$
limit by a phenomenological fit, $D_L=D_\infty+bL_n^{-2}$, obtaining
$D_\infty=0.0012(1)l^2$/MCS and $b=0.0433(1)l^4$/MCS.
This extrapolated value, which does not depend appreciably on the particular
algebraic form used, can be translated into the real,
physical diffusion constant $D=0.0012(1) l^2 \nu \exp(-\epsilon/k_B T)$.
In principle, values of the attempt frequency $\nu$ and the barrier energy
$\epsilon$ could be determined in a realistic quasicrystal model,
for example, by using a molecular dynamic simulation.

In order to set a scale, the prefactor 0.0012(1) of the diffusion
constant we obtained, can be compared with the analogous prefactor for
vacancy-assisted self-diffusion, which is typically $\sim10^{-4}$ at
room temperature.  Generally, since the vacancy prefactor is
essentially a Boltzmann factor associated with the vacancy activation
energy, the difference between the prefactors should be most pronounced
in real quasicrystals at temperatures at which the phason activation
begins to saturate.  For some quasicrystalline materials these
temperatures are estimated at above 1000K, where the prefactors are
comparable.\cite{phas} However, because of the exponential dependence
of the diffusion constants on activation energies, it is difficult to
estimate them reliably for real quasicrystals.

A further complication may arise if the phason flip in a real quasicrystal
involves a rearrangement of a larger group of atoms.
This would not only limit the accuracy of our estimates, but in such case
the phason flips might have to be accompanied with a vacancy motion
to effect diffusion.
The vacancy motion may also play an important role in facilitating
phason flips, as suggested by a molecular dynamics simulation
of a two dimensional, two component Lennard-Jones quasicrystal,
which showed that vacancy motion in that system was strongly
correlated with phason flips.\cite{marco}

Since the phason Goldstone mode requires correlated flips of
infinitely many rhombic dodecahedra arranged in families of continuous
sheets, isolated phason flips would generically require a non-zero
energy.
However, as pointed out in \cite{kk}, this would have a negligible
effect in an equilibrated quasicrystal at sufficiently high
temperature where a random tiling description becomes appropriate.
On the other hand, this implies that the new mechanism of self-diffusion
is, at a sufficiently high temperature, not so much a consequence of
the existence of the phason Goldstone mode, i.e. of incommensurability,
but of a particular tiling character of the structure.
Therefore, the same mechanism should also be effective at a sufficiently
high temperature in periodic crystal approximants of the quasicrystal.
In fact, the diffusion constants $D_2=0.0010(1)l^2$/MCS and
$D_3=0.0011(1)l^2$/MCS that we determined for the infinite
($m\rightarrow\infty$) periodic approximants with $n$=2 and 3,
although slightly {\it smaller} than what we found for the Ammann tiling,
are consistent with $D_\infty=\lim_{n\rightarrow\infty}D_n$.

It is interesting to examine if there is a simple scaling
form that describes all of our data.
Since the short time behavior
that we observed is independent of the system size,
the scaling, if it exists, should take the form
$R_L^2=\tau_L^\beta g(t/\tau_L)$,
where $\tau_L$ is a characteristic diffusion time that depends on the system
size $L$, and the scaling function $g$ should have the asymptotic form
$ax^\beta$ for $x\ll 1$.
On the other hand, the normal diffusion in the long time limit demands
that $g(x)=dx+c$, for $x\gg 1$.
This, in turn, implies that $D_L^{\beta/(1-\beta)} C_L$
should be independent of $L$, where $D_L$ and $C_L$ are
determined from the long time behavior Eq.~(\ref{normald}).
Our measured values are inconsistent
with this conclusion since $C_L$ changes sign as $L$ is increased.
It is possible that our systems are not sufficiently large and
the discrepancy is due to the corrections to scaling.
Another possible explanation for the apparent lack of a simple finite-size
scaling is that there are two mechanisms,
or two distinct length scales, responsible for the diffusion.
Using the six-dimensional representation,
where the tiling is a fluctuating three
dimensional membrane,\cite{elser}
it is tempting to associate the two mechanisms with
the decomposition of the membrane fluctuations into the relative and
the center of mass motions.
Indeed, it is the relative motion of the membrane that is responsible
for permutations of vertices.
In the physical space, such motions can be viewed as a cloud of
phason flips that must accompany the permutation motion of a vertex.
The size of this cloud would be finite,
on the order of unity for Ammann tilings, and independent of $L$
in the scaling limit.
Similarly, the characteristic time for the short time correlated
motion of vertices in the cloud, at which the short time subdiffusive
transport $t^\beta$ crosses over to the normal diffusion,
would approach a finite value $\tau_\infty=(a/D_\infty)^{1/(1-\beta)}$
in the scaling limit $L\rightarrow\infty$.
These conclusions are consistent with $D_L$ approaching a finite,
non-zero value $D_\infty$ in this limit and with the characteristic
time $\tau_\infty=1.9(4)~10^3$MCS comparable to the average time needed
to permute vertices inside the rhombic icosahedron in Fig. 1c.

Alternatively, the discrepancy could arise from anomalous corrections to
the normal diffusion [$Ct^{1\over\mu}$, with $1<\mu<2$,
$C\sqrt{t\ln t}$, or $C\sqrt{t}$,
instead of $C$ in Eq. (\ref{normald})]
due to any long-time correlations.\cite{reports}
It is also difficult to rigorously rule out a possibility
that the long time transport is anomalous
with $R^2\sim Dt/\ln t$, or $R^2\sim Dt^\mu$ and
$\mu$ very close to one.\cite{mon}
The waiting time distribution provides a valuable tool for
further clarification of these possibilities,
and our preliminary analysis shows that the distribution could
be consistent with a $\mu$ close to 2.

We have demonstrated that phason dynamics does indeed result in
self-diffusion of vertices of the icosahedral Ammann tiling, in
agreement with a general scenario conjectured for quasicrystals.\cite{kk}
On short time scales, we found that the transport is subdiffusive
with the exponent $\beta\approx0.57(1)$.
We determined the diffusion constant to be
$D=0.0012(1) l^2 \nu\exp(-{\epsilon\over k_B T})$, and we concluded
that,
in a real quasicrystal, it could be an order of magnitude
larger than the room temperature vacancy-assisted diffusion constant
with comparable parameters.
The two mechanisms are expected to enhance each other, especially at
lower temperatures
where vacancies would be more effective in randomizing the phason field.
However, a study of the low-temperature diffusion in energetically
stabilized Ammann tilings or its approximants, is a significantly more complex
problem which will have to be studied separately.
Such study should also shed a light on the pinning-depinning transition
in phason dynamics.
If the phason-assisted self-diffusion is significant in a quasicrystal,
then a strong change in the diffusion constant should be experimentally
observed at its depinning transition.
Additional work is also needed to resolve the questions about the
finite size scaling for the phason-assisted diffusion.


We are grateful to P. Kalugin for communicating his results prior to
publication,\cite{kprivate} to T. Soma for generous help with Fig. 1, and to
P. Young and, especially, J. Deutsch for illuminating
discussions and hospitality at the University of California Santa Cruz
where a substantial part of this work was completed.
This research was supported in part by the NSF grant No. DMR9215231.

\begin{figure}
\label{fig1}
\caption{(a) The six icosahedral five-fold symmetry
directions of the Ammann tilings
(arrows) and the two types of Ammann rhombohedra
with edges along these directions.
(b) Phason flip in the Ammann tiling as a
rearrangement of a rhombic dodecahedron.
(c) The vertices inside a rhombic
icosahedron (emphasized by larger spheres)
that are separated by a short rhombus
diagonal can be permuted as a result of ten
successive phason flips.}
\end{figure}
\begin{figure}
\caption{Log-log plot of the average square vertex displacement versus
time, for several crystal approximants.
The inset shows extrapolation to the infinite Ammann tiling limit of
the resulting diffusion constants $D_L$.}
\label{fig2}
\end{figure}
\end{document}